\documentclass[aps,amssymb,amsmath,prl,reprint,noshowpacs,]{revtex4-1}
\usepackage{times}
\usepackage{amssymb,amsmath,graphicx}
\usepackage[usenames]{color}
\usepackage{bbold}

\newcommand{\vect}[1]{\boldsymbol{#1}}

\newcommand{\iunit}{\text{i}}
\usepackage{textcomp}
\newcommand{\murm}{\hbox{\textmu}}

\newcommand{\figwidth}{0.45\textwidth} 

\begin{document}
\title{Superemitters in Hybrid Photonic Systems: A Simple Lumping Rule for \\the Local Density of Optical States and its Break-Down at the Unitary Limit}
\author{Martin Frimmer}
\affiliation{Center for Nanophotonics, FOM Institute AMOLF,
Science Park 104, 1098 XG Amsterdam, The Netherlands}
\email{frimmer@amolf.nl}
\author{A. Femius Koenderink}
\affiliation{Center for Nanophotonics, FOM Institute AMOLF,
Science Park 104, 1098 XG Amsterdam, The Netherlands}








\begin{abstract}
We theoretically investigate how the enhancement of the radiative decay rate of a spontaneous emitter provided by coupling to an optical antenna is modified when this ``superemitter" is introduced into a complex photonic environment that provides an enhanced local density of optical states (LDOS) itself, such as a micro-cavity. We show that photonic environments with increased LDOS further boost the performance of antennas that scatter weakly, i.e. that are far from the unitary limit, for which a simple multiplicative LDOS lumping rule holds. In contrast, enhancements provided by antennas close to the unitary limit, i.e. antennas close to the limit of maximally possible scattering strength, are strongly reduced by an enhanced LDOS of the environment. Thus, we identify multiple scattering in hybrid photonic systems as a powerful mechanism for LDOS engineering.
\end{abstract}
\date\today

\maketitle


Optics encompasses the most fascinating part of the electromagnetic spectrum, due to its {energetic} overlap with electronic transitions in matter. Nano-photonics aims at controlling such transitions by  molding light at sub-wavelength scales. Purcell first predicted that resonators modify the radiative lifetime of spontaneous emitters, a property held to be intrinsic to the source until then~\cite{Purcell1946}. Modern literature discusses the Purcell effect in terms of the local density of optical states (LDOS), a fundamental quantity governing spontaneous emission, thermal radiation, and vacuum forces~\cite{Novotny2006}. Two tools have emerged to shape the LDOS: on the one hand, interfaces~\cite{Drexhage1970,*Snoeks1995}, photonic crystals~\cite{Lodahl2004,Barth2009,Sar2011} and dielectric micro-cavities~\cite{Vahala2005} modulate the LDOS on length scales of order $\lambda/2$ via interference. On the other hand, optical antennas \cite{Anger2006,*Kuehn2006,Schietinger2009,*Kinkhabwala2009} employ intrinsic plasmonic material resonances to enhance the LDOS on $\lambda/20$ length scales. Optical antennas are so small compared to the wavelength $\lambda$ that a source-antenna ensemble essentially radiates as a dipole. This similarity to a bare molecule, but with much higher radiative rate, has prompted the term ``superemitter"~\cite{Farahani2005} to refer to this source-antenna entity.
The availability of photonic building blocks on such different length scales raises the exciting idea of integrating deep-subwavelength superemitters in much larger dielectric structures to obtain a combined advantage, e.g. by  placing a nano-antenna inside a micro-resonator~\cite{Benson2011,Xiao2012}.
In view of these recent developments it is imperative to understand how the LDOS of such hybrid systems emerges from that of the separate entities.
Since nano-optic devices can be interpreted as lumped optical elements~\cite{Engheta2005,*Engheta2007,*Al`u2008,*Al`u2008b,Greffet2010}, where the LDOS acts as an impedance for spontaneous emitters, much like the resistivity experienced by a classical AC current source due to the fact that it radiates energy into the far field, some simple circuit rule might be hoped for, which lumps the LDOS provided individually by the photonic building blocks.

This Letter theoretically investigates how the decay rate enhancement provided by a superemitter varies when it is positioned within a larger photonic system, i.e., how the LDOS lumps. We consider two archetypical examples: First, we investigate a superemitter coupled to high-Q resonances. Second, we consider a superemitter in front of a perfect mirror. Our analytic, yet fully electrodynamic model yields a simple multiplicative LDOS lumping rule for moderate antenna factors.
For antennas operating close to the unitary limit, however, this simple lumping rule breaks down and the total enhancement becomes proportional to the \emph{inverse} LDOS of the background system.
This insight paves the way for  engineering the LDOS by exploiting multiple scattering in hybrid photonic systems.

The radiative decay rate enhancement of a spontaneous quantum emitter can be calculated via the power required to sustain a constant classical point current $\vect{j}=\dot{\vect{p}}_0$ that loses energy by radiation~\cite{Novotny2006}.
This power equals the cycle averaged work per unit time done by the source's electric field on its own dipole moment $\vect{p}_0$. The electric field generated at position $\vect{r}$ due to a source $\vect{p}_0$ at $\vect{r_0}$ is calculated via the electric Green function $\vect{G}(\vect{r},\vect{r_0})$ of the respective system.
This yields the  power required to drive the source $P=\frac{1}{2}\omega\,\vect{p}^T_0\text{Im}\vect{G}(\vect{r}_0,\vect{r}_0)\vect{p}_0$. For lossy environments, this expression describes the total decay rate, i.e., radiation plus quenching induced by the environment.
We use the term LDOS to refer to Im$\vect{G}$  projected on a unit vector along $\vect{p}_0$, i.e. to the decay rate modification of a molecular dipole $\vect{p}_0$ at position $\vect{r}_0$.
Every strategy to boost decay rates via a photonic structure, be it large and dielectric, or a nano-antenna, represents a modification of Im$\vect{G}$.
The small size and dipolar nature of a nano-antenna however suggest to interpret its rate enhancement rather as an enhancement of the dipole moment $\vect{p}_0$ instead of a change in Im$\vect{G}$~\cite{Farahani2005}.
The simplest optical antenna is just a particle with polarizability tensor $\boldsymbol{\alpha}(\omega)$~\cite{Kuehn2006,Anger2006}. At distance $d$ in the near field of an emitter, the particle  acquires a large dipole moment $\propto{\boldsymbol{\alpha}}/{d^3}$. The total dipole moment of the emitter-particle assembly $\vect{p}=[1 + \boldsymbol{\alpha}/d^3 ]\vect{p_0}$ can exceed by far the source's dipole moment $\vect{p}_0$, rationalizing the term `superemitter'. If $d\ll\lambda$, the power radiated by the superemitter~\footnote{We assume the source to be aligned radially to an isotropic particle. For a full vectorial expression see Ref.~\onlinecite{Sipe1974}.}
\begin{equation}
\label{eq:LarmorSuper}
P=\dfrac{1}{2}\omega\left|\vect{p}_0\right|^2\,\text{LDOS}_\text{B}(\vect{r}_0)\times A
\end{equation}
exceeds that of the bare source by the antenna factor $A=\left|1+{\alpha}/d^3\right|^2\approx\left|{\alpha}\right|^2/d^6$. The rate enhancement provided by the embedding background system at the location of the superemitter is described by LDOS$_\text{B}$. Mie calculations show that $A$ accurately describes antenna particles up to 60\,nm in diameter in vacuum~\cite{Mertens2007}.
Equation~\eqref{eq:LarmorSuper} suggests that a superemitter can be used as replacement of a bare molecule to probe the LDOS of a larger photonic system, since in this reasoning a simple product rule lumps the enhancements provided by the antenna $A$ and by the photonic environment's LDOS$_\text{B}$.

We analyze the hypothesis of a simple lumping rule in an analytic electrodynamic point scattering theory, which is exact to all multiple scattering orders, with the sole assumption that the scatterers that constitute the antenna are well approximated as point dipoles. The scatterers $1\ldots N$ acquire dipole moments $\vect{p}_1\ldots\vect{p}_N $ in proportion to their polarizabilities $\boldsymbol{\alpha}_1\ldots\boldsymbol{\alpha}_N$, and the electric fields $\vect{E}(\vect{r}_1)\ldots\vect{E}(\vect{r}_N)$ at their locations $\vect{r}_n$, according to the linear self-consistent set of equations~\cite{Weber2004,GarciadeAbajo2007}
\begin{equation}
\vect{p}_n=\boldsymbol{\alpha}_n\left[
\vect{E}_\text{in}(\vect{r}_n) + \sum_{m\neq n}
\vect{G}_\text{B}(\vect{r}_n,\vect{r}_m)\cdot \vect{p}_m \right].
\label{eq:selfconsistent}
\end{equation}
By construction, the antenna described by the $\boldsymbol\alpha_n$ is explicitly separated from the background system that it probes, which is quantified by its Green function $\vect{G}_\text{B}$.
For a consistent theory, three facts need to be taken into account. First, the particle polarizability directly depends on the background via~\cite{Greffet2010}
\begin{equation}
\boldsymbol{\alpha}^{-1}_{n}=\boldsymbol{\alpha}^{-1}_{n,\text{0}}
- \vect{G}_B(\vect{r}_n,\vect{r}_n),
\label{eq:alphacorrection}
\end{equation}
where $\boldsymbol{\alpha}_{n,\text{0}}$ is the electrostatic polarizability.
Note that isotropic particles can \emph{acquire} an anisotropy due to the anisotropy in radiation damping given by Im$\vect{G}_B$ and resonance energy shift due to Re$\vect{G}_B$ in complex photonic systems~\footnote{Since $\text{Re}\vect{G}_{\text{vac}}$ diverges at the source, and one is only interested in relative frequency shifts, it is commonly included in $\vect{\alpha}_0$ to yield a finite resonance frequency $\omega_0$.}.
Second, the source in our model is a single dipole $\vect{p}_0$ at $\vect{r}_0$ (with $\boldsymbol{\alpha}_{0}=0$) so $\vect{E}_{\text{in}}(\vect{r}_n)=\vect{G}_\text{B}(\vect{r}_n,\vect{r}_0)\vect{p}_0$.
Third, the total decay rate of the source in the full system is found via the cycle-averaged work
done by the \emph{total} electric field on the source $\vect{p}_0$~\cite{Novotny2006}.
Therefore, using a source of unit strength $\vect{p}_0$, the LDOS equals
\begin{equation}
  \vect{p}^T_0\cdot\text{Im}
\vect{G}_{B}(\vect{r}_0,\vect{r}_0)\cdot  \vect{p}_0 +
\text{Im} \sum_{n\geq 1} \vect{p}_0^T \cdot
  \vect{G}_B(\vect{r}_0,\vect{r}_n) \cdot \vect{p}_n.
 \label{eq:totalldos}
\end{equation}
The first term is the LDOS$_\text{B}$ provided by the background system itself in absence of the antenna, while the second term is the contribution from the antenna.
We calculate the LDOS of the hybrid system by evaluating Eq.~\eqref{eq:totalldos}, after solving the $3N$ linear equations in Eq.~\eqref{eq:selfconsistent} for the induced dipole moments $\vect{p}_1\ldots\vect{p}_N$.
We use the exact Green function for a sphere~\cite{Tai1994} and a planar interface~\cite{Novotny2006,Paulus2000} to evaluate how a superemitter probes the two canonical cases of micro-cavity resonances and non-resonant interfaces~\cite{Drexhage1970}, respectively.

\begin{figure}
\includegraphics[width=\figwidth]{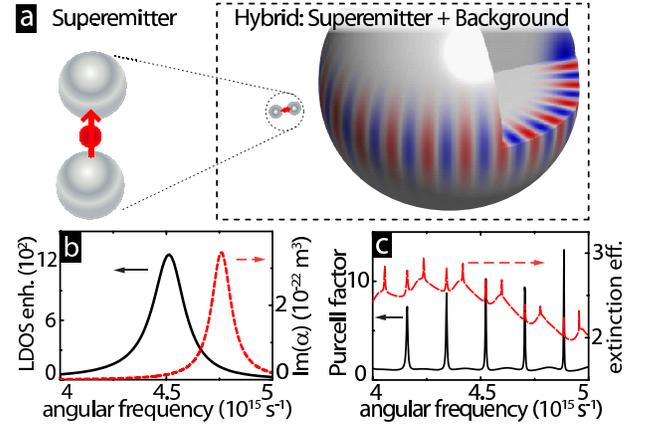}
\caption{(a) Left: Sketch of superemitter formed by two silver spheres. The fluorescent source is located between the two particles with its dipole moment along the symmetry axis. Right: Hybrid photonic system comprised of superemitter embedded in background system formed by dielectric sphere. Red and blue colors illustrate fields of a whispering gallery mode. (b) Dashed line: polarizability of a single antenna particle in vacuum. Solid line: antenna enhancement factor for superemitter sketched in (a) in vacuum. (c) Dashed line: Extinction efficiency of dielectric sphere showing narrow Mie-resonances. Solid line: Purcell factor of the Mie-sphere 50\,nm from its surface.}
\label{Fig:bareSuperemitter}
\end{figure}
As a superemitter we consider a source in the gap of a dimer antenna with its dipole $\vect{p}_0$ along the symmetry axis [see Fig.\ref{Fig:bareSuperemitter}(a)]. The Ag spheres forming the dimer, each of diameter 40\,nm, have a center-center distance of 60\,nm~\footnote{For spherical particles of volume $V$ and dielectric constant $\epsilon(\omega)$, the electrostatic polarizability is simply $3V(\epsilon-1)/(\epsilon+2)$. A Drude model for $\epsilon$ yields a Lorentzian. We choose $\omega_0$=$4.76\times10^{15}\,\text{s}^{-1}$ and $\gamma$=$8.3\times10^{12}\,\text{s}^{-1}$ to model the Ag particles.}.
The dashed line in Fig.~\ref{Fig:bareSuperemitter}(b) shows Im$\boldsymbol{\alpha}$ of a single antenna particle in vacuum. The solid line is the decay rate enhancement of the source in the antenna gap with the antenna in vacuum. The antenna factor reaches about 1200, in good agreement with full multipole calculations~\cite{Koenderink2010}, which justifies the use of a dipole model. The dimer  resonance is red-shifted and broadened by super-radiant damping compared to the single particle due to longitudinal symmetric dipolar plasmon hybridization~\cite{Prodan2003}.

As a background system we now consider a glass sphere (n=1.5) with diameter 2.4\,\murm m, supporting whispering gallery modes, illustrated in Fig.~\ref{Fig:bareSuperemitter}(a). The dashed line in Fig.~\ref{Fig:bareSuperemitter}(c) is the sphere's extinction efficiency~\cite{Bohren1983}. Characteristic resonances are clearly visible as sharp peaks with $Q\approx10^2$. The Purcell factor for a radially oriented source 50\,nm from the sphere surface [solid line in Fig.~\ref{Fig:bareSuperemitter}(c)] reaches moderate values around 10. Only every second peak in extinction appears in the Purcell enhancement, which reflects the field orientation of the modes according to the common TE/TM type classification~\cite{Agha2006}.
We now place the superemitter with the center of the closest antenna particle 50\,nm from the sphere's surface and the symmetry axis pointing radially outwards [see Fig.~\ref{Fig:bareSuperemitter}(a)].
\begin{figure}
\includegraphics[width=\figwidth]{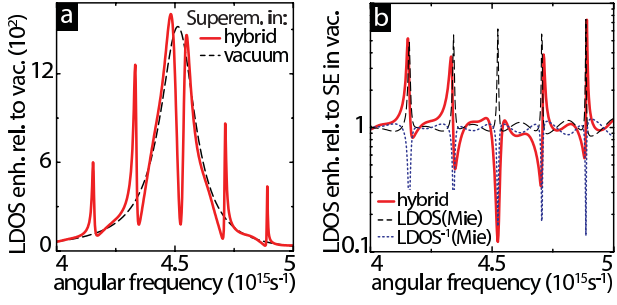}
\caption{(a) Solid line (hybrid system): Decay rate of source in the gap of the superemitter next to the Mie-sphere normalized to rate of source in vacuum.
Dashed line (superemitter in vacuum): Decay rate of source in the gap of the superemitter in vacuum normalized to rate of source in vacuum. At the Mie-resonances, the hybrid LDOS enhancement is  drastically modified compared to that of the antenna in vacuum. (b) Solid line (hybrid system): Decay rate enhancement in hybrid system [solid line in (a)], normalized to  the rate enhancement in the superemitter in vacuum [dashed line in (a)]. Off antenna resonance, the superemitter benefits from the LDOS enhancement offered by the sphere (dashed line), while on antenna resonance the enhancement is suppressed by the \emph{inverse} of the sphere's LDOS (dotted line).}
\label{fig:SphereData}
\end{figure}
The solid line in Fig.~\ref{fig:SphereData}(a) is the decay rate enhancement experienced by the emitter embedded in the dimer antenna which in turn is located next to the Mie-sphere, i.e. the rate in the hybrid system normalized to the rate of the bare source in vacuum.
While the overall shape of the enhancement provided by the antenna in vacuum [Fig.~\ref{fig:SphereData}(a), dashed] is still visible, sharp features appear at five spectral positions coinciding with the sphere's whispering gallery modes [cf. Fig.~\ref{Fig:bareSuperemitter}(c)].
To illustrate the effect of the background system on the superemitter we now normalize the decay rate enhancement provided by the hybrid system [solid line in Fig.~\ref{fig:SphereData}(a)] to the enhancement provided by the bare antenna in vacuum [dashed line in Fig.~\ref{fig:SphereData}(a)] and plot it as the solid line in Fig.~\ref{fig:SphereData}(b).
The sharp enhancements located in the wings of the antenna resonance follow the LDOS of the sphere, denoted by the dashed line in Fig.~\ref{fig:SphereData}(b). Therefore, off antenna resonance, at still significant antenna factors, we indeed find the anticipated behavior of a superemitter, i.e., its already enhanced decay rate compared to the bare molecule is further boosted by the presence of a high-Q resonance of the background.
Furthermore, we note the dispersive features in the enhancement, which swap orientation upon crossing the antenna resonance.
Surprisingly, however, on antenna resonance, the LDOS enhancement is strongly suppressed by the Mie-sphere. This LDOS suppression close to antenna resonance cannot be explained by a spoiling or detuning of the cavity by the antenna, since this would only result in a shift or absence of a sharp line of extra enhancement on top of the bare antenna factor.
Also,  Waldron's formula predicts that our antenna does not significantly shift or spoil the micro-sphere resonances due to their large mode volumes compared to $\boldsymbol{\alpha}$ and their low $Q$-factors~\cite{Agha2006,Koenderink2005a}. Figure~\ref{fig:SphereData}(b) hence implies a  spoiling of the antenna by a cavity resonance tuned close  to the antenna resonance, instead of vice versa.

\begin{figure}
\includegraphics[width=\figwidth]{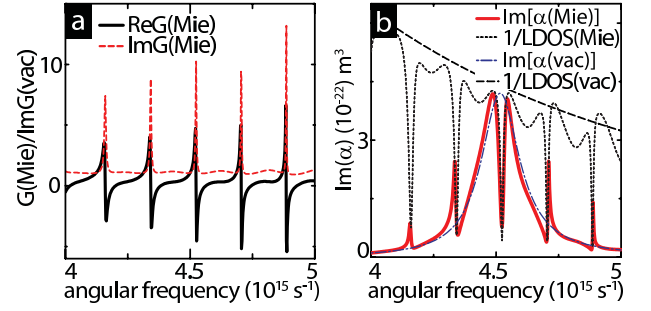}
\caption{(a) Real (solid) and imaginary (dashed) parts of the radial component of the Green function of the Mie-sphere 50\,nm from sphere's surface. These terms enter the radiation correction to $\boldsymbol{\alpha}_\text{Mie}$. (b) Solid line: Radial component of the antenna polarizability $\boldsymbol{\alpha}_\text{Mie}$ when located 50\,nm from the Mie-sphere. Dash dotted line: $\boldsymbol{\alpha}_\text{vac}$ for same antenna in vacuum. While $\boldsymbol{\alpha}_\text{vac}$ is limited by the inverse vacuum LDOS (dashed line), $\vect{\alpha}_\text{Mie}$ is bounded by the inverse of the sphere's LDOS (dotted line), which leads to the suppression of $\vect{\alpha}_\text{Mie}$ close to the antenna resonance.}
\label{fig:Alpha}
\end{figure}
In order to understand the truly surprising spoiling of the antenna enhancement by a large background LDOS, e.g. of a micro-cavity, one needs to interpret Eq.~\eqref{eq:LarmorSuper} correctly, by taking the proper radiation damping correction according to Eq.~\eqref{eq:alphacorrection} into account.
For simplicity we now discuss an antenna significantly smaller than the wavelength and described as a single scatterer with polarizability $\boldsymbol\alpha$~\footnote{It is easy to show that a small dimer antenna is described as a single scatterer with $\boldsymbol{\alpha}_\text{eff}$=$2\left[\boldsymbol{\alpha}_\text{B}^{-1}-\vect{G}_\text{B}(\vect{r}_1,\vect{r}_2)\right]^{-1}$}.
Equation~\eqref{eq:alphacorrection} ensures that the optical theorem $\text{Im}{\alpha}\ge \text{LDOS}_\text{B}\,\left|{\alpha}\right|^2$
is fulfilled, where equality holds for the case of no material loss~\footnote{For a full vectorial expression of the optical theorem see Ref.~\onlinecite{Sipe1974}.}.
Thereby Eq.~\eqref{eq:alphacorrection} strictly bounds the polarizability according to
\begin{equation}
\label{eq:UnitaryLimit}
\text{Im}(\boldsymbol{\alpha})\le \frac{1}{\text{Im}{\vect{G}}_{\text{B}}},
\end{equation}
which is a general form of the unitary limit, a  fundamental relation in any scattering theory~\cite{Newton2002}.
In vacuum, this limit is well-known for extinction cross section as $\sigma_{\text{ext}}= 4\pi k\text{Im}\alpha\le\frac{3}{2\pi}\lambda^2$, a limit reached by an ideal scatterer on resonance, and closely approached by any plasmon particle above 20\,nm in size~\cite{Mertens2007}.
The  unitary limit Eq.~\eqref{eq:UnitaryLimit}  states  that $\boldsymbol{\alpha}$ of a strong scatterer, and hence the dipole moment it acquires, is proportional to the \emph{inverse} LDOS.
Intuitively, since a strong scatterer is predominantly damped by radiation, increasing the background LDOS further increases its loss and therefore suppresses the scatterers response, which reflects in a reduced polarizability with increased spectral width~\cite{Buchler2005}.

To quantitatively verify that the unitary limit indeed governs the hybrid system's LDOS
we evaluate Eq.~\eqref{eq:totalldos} for a physically small superemitter. The hybrid system's LDOS is then dominated by $\text{Re}\vect{G}_{\text{vac}}$ to read $\text{Re}\vect{G}_{\text{vac}}(\vect{r}_0,\vect{r}_1)\text{Im}\boldsymbol{\alpha}\text{Re}\vect{G}_{\text{vac}}(\vect{r}_1,\vect{r}_0)$, which is of order $1/d^6$ (with $d=|\vect{r}_1-\vect{r}_0|$) and precisely yields Eq.~\eqref{eq:LarmorSuper}~\footnote{Note that $\vect{G}_{\text{vac}}\propto1/d^3$ while $\vect{G}_{\text{B}}$ is of order $k^3$.}.
To illustrate the effect of the background system on Im$\vect{\alpha}_\text{Mie}$ of a particle located in close vicinity of the sphere we plot in Fig.~\ref{fig:Alpha}(a) the radial component of the sphere's Green function $\vect{G}_\text{Mie}$ 50\,nm from the surface.
Real and imaginary part of $\vect{G}_{\text{Mie}}$ show the typical dispersive and dissipative line-shape of a resonance, respectively. Naturally, Im$\vect{G}_\text{Mie}$ equals the micro-cavity Purcell factor at the source position [cf. Fig.~\ref{Fig:bareSuperemitter}(c)].
In Fig.~\ref{fig:Alpha}(b) we plot as the solid line the radial component of Im$\vect{\alpha}_\text{Mie}$ of an antenna located 50\,nm from the sphere surface. Note that the values of $\vect{G}_\text{Mie}$ in Fig.~\ref{fig:Alpha}(a) are the correction terms entering Eq.~\eqref{eq:alphacorrection} causing the modification of $\vect{\alpha}_\text{Mie}$ close to the sphere [solid line in Fig.~\ref{fig:Alpha}(b)] with respect to $\vect{\alpha}_\text{vac}$ in vacuum [dash-dotted line in Fig.~\ref{fig:Alpha}(b)].
Close to antenna resonance Im$\boldsymbol{\alpha}_\text{Mie}$ is indeed limited by the inverse of the sphere's LDOS [Fig.~\ref{fig:Alpha}(b), dotted line].
Off antenna resonance, the correction to the broad $\vect{\alpha}_\text{Mie}$ due to the narrow Re$\vect{G}_{\text{Mie}}$ creates characteristic Fano-resonances~\cite{Frimmer2012}.
The transition from enhancement to inhibition in the lumped LDOS  between the limits of weak and maximally strong scattering is captured by carrying out the correction according to Eq.~\eqref{eq:alphacorrection} in the antenna factor $A$ in Eq.~\eqref{eq:LarmorSuper}, such that when neglecting the real frequency shift the radiated power reads $P\propto\left| \vect{\alpha}^{-1}_{\text{0}}-\iunit\,\text{LDOS}_\text{B}\right|^{-2}\text{LDOS}_\text{B}$.
For small $\vect{\alpha}_\text{0}$, i.e. in the limit of weak scattering, $P\propto\left|\vect{\alpha}_\text{0}\right|^2\text{LDOS}_\text{B}$ since weak Rayleigh scatterers \emph{are} constant current sources unaffected by the unitary limit~\cite{Motsch2010}. Therefore, for weak scatterers a simple multiplicative lumping rule for $A$ and LDOS$_\text{B}$ \emph{does} hold. In the limit of strong scattering $P\propto\text{LDOS}_\text{B}^{-1}$, since a scatterer at the unitary limit is \emph{not} a constant current source~\cite{Kalkbrenner2005,Castanie2011}. Accordingly, close to antenna resonance, the enhancement curve of the hybrid system [solid line in Fig.~\ref{fig:SphereData}(b)] follows the \emph{inverse} of the Mie-sphere's LDOS [dotted line in Fig.~\ref{fig:SphereData}(b)].

To illustrate that our results are generally valid beyond the specific case of high-Q Mie resonances, we examine a near-perfect mirror ($\epsilon$=-200) as a background system that modifies the LDOS without any resonances~\cite{Drexhage1970}.
\begin{figure}
\includegraphics[width=\figwidth]{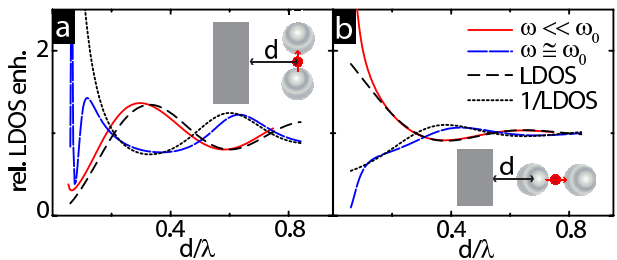}
\caption{Relative LDOS enhancement for superemitter as a function of distance to a near-perfect mirror. (a) Superemitter oriented parallel to mirror surface [see inset]. Solid line: Superemitter with source far below antenna rensonance [$\omega=4.0\times10^{15}\,\text{s}^{-1}$, cf. Fig.~\ref{Fig:bareSuperemitter}(b)], traces the LDOS of the mirror [dashed line], except at small superemitter-mirror separations. A source close to antenna resonance [long dashed line, $\omega=4.5\times10^{15}\,\text{s}^{-1}$] traces the inverse of the mirror LDOS [dotted line], except at small separations. (b) Same as (a) for superemitter perpendicular to mirror [see inset].}
\label{fig:MirrorData}
\end{figure}
In Fig.~\ref{fig:MirrorData}(a), we plot the decay rate of the superemitter in front of the mirror normalized to the superemitter in vacuum as a function of distance to the mirror. The antenna axis is parallel to the mirror [see sketch in inset]. Far below antenna resonance, the enhancement [Fig.~\ref{fig:MirrorData}(a), solid line] follows the mirror LDOS, plotted as the dashed line, as expected from the multiplicative LDOS lumping rule.
Close to antenna resonance [long dashed line], the antenna is close to the unitary limit and therefore the total rate enhancement follows the inverse of the mirror LDOS, indicated by the dotted line. At intermediate frequencies we observe a smooth transition between the two cases illustrated in Fig.~\ref{fig:MirrorData} (data not shown). For the superemitter oriented perpendicularly to the mirror [Fig.~\ref{fig:MirrorData}(b)] we observe the analog behavior as for the parallel case. The product lumping rule and its cross-over to \emph{inverse} proportionality are hence generic. The framework  only breaks down for very small separations between superemitter and mirror, where the superemitter size is comparable to its distance to its own mirror image.  Here superemitters also sense  gradients in $\vect{G}$, which may provide a tailorable analogon to recent experiments on mesoscopic quantum dots probing multipolar LDOS~\cite{Andersen2011}.

In conclusion, we have examined how the LDOS inside a superemitter probes the LDOS of a complex photonic environment. Generally for any superemitter with a moderate antenna factor the LDOS enhancements of  antenna and background multiply and a small superemitter can therefore serve as an LDOS probe for a large background system, exactly as the term suggests~\cite{Farahani2005}. In surprising contrast, a superemitter with an antenna at the unitary limit probes the \emph{inverse} background LDOS, since increasing radiation damping reduces the polarizability of strong scatterers~\cite{Newton2002}.
Our findings imply that if a \emph{general} lumping rule for optical source impedances~\cite{Greffet2010}, i.e. LDOS, can be found at all, it must take into account not just the bare superemitter LDOS, but also how close the antenna is to the unitary limit, which in turn depends on the background system.
Our insights bear a plethora of exciting prospects for engineering of LDOS with the toolbox of multiple scattering~\cite{Verslegers2012}. One might consider nano-manipulative switching of superemitters by moving them with respect to high-Q resonators~\cite{Mazzei2007}.
Furthermore, our work could lead the quest for ultra-strong optical antennas towards counterintuitive hybrids of nano-antennas embedded in photonic bandgap devices.
As a further implication, our findings shed new light on attempts to use a scatterer as a broadband probe of thermally populated modes, since the scatterer \emph{acquires} spectral features thanks to the environment that is to be probed~\cite{Wilde2006,*Liu2011,*Lin2003,*Han2010}.
Notably, hybrid photonic systems composed of superemitters operated off resonance together with a micro-resonator might benefit
from large field enhancements inside the antenna that occur with convenient moderate-Q cavities right at the Fano resonances in hybrid LDOS, an exciting prospect for single molecule detection~\cite{Vollmer2008}.

\begin{acknowledgments}
This work is part of the research program of the ``Stichting voor Fundamenteel
Onderzoek der Materie (FOM)'', which is financially supported by the
``Nederlandse Organisatie voor Wetenschappelijk Onderzoek (NWO)''.
\end{acknowledgments}

\bibliography{Manuscript_Frimmer}

\begin{thebibliography}{51}%
\makeatletter
\providecommand \@ifxundefined [1]{%
 \@ifx{#1\undefined}
}%
\providecommand \@ifnum [1]{%
 \ifnum #1\expandafter \@firstoftwo
 \else \expandafter \@secondoftwo
 \fi
}%
\providecommand \@ifx [1]{%
 \ifx #1\expandafter \@firstoftwo
 \else \expandafter \@secondoftwo
 \fi
}%
\providecommand \natexlab [1]{#1}%
\providecommand \enquote  [1]{``#1''}%
\providecommand \bibnamefont  [1]{#1}%
\providecommand \bibfnamefont [1]{#1}%
\providecommand \citenamefont [1]{#1}%
\providecommand \href@noop [0]{\@secondoftwo}%
\providecommand \href [0]{\begingroup \@sanitize@url \@href}%
\providecommand \@href[1]{\@@startlink{#1}\@@href}%
\providecommand \@@href[1]{\endgroup#1\@@endlink}%
\providecommand \@sanitize@url [0]{\catcode `\\12\catcode `\$12\catcode
  `\&12\catcode `\#12\catcode `\^12\catcode `\_12\catcode `\%12\relax}%
\providecommand \@@startlink[1]{}%
\providecommand \@@endlink[0]{}%
\providecommand \url  [0]{\begingroup\@sanitize@url \@url }%
\providecommand \@url [1]{\endgroup\@href {#1}{\urlprefix }}%
\providecommand \urlprefix  [0]{URL }%
\providecommand \Eprint [0]{\href }%
\providecommand \doibase [0]{http://dx.doi.org/}%
\providecommand \selectlanguage [0]{\@gobble}%
\providecommand \bibinfo  [0]{\@secondoftwo}%
\providecommand \bibfield  [0]{\@secondoftwo}%
\providecommand \translation [1]{[#1]}%
\providecommand \BibitemOpen [0]{}%
\providecommand \bibitemStop [0]{}%
\providecommand \bibitemNoStop [0]{.\EOS\space}%
\providecommand \EOS [0]{\spacefactor3000\relax}%
\providecommand \BibitemShut  [1]{\csname bibitem#1\endcsname}%
\let\auto@bib@innerbib\@empty
\bibitem [{\citenamefont {Purcell}(1946)}]{Purcell1946}%
  \BibitemOpen
  \bibfield  {author} {\bibinfo {author} {\bibfnamefont {E.~M.}\ \bibnamefont
  {Purcell}},\ }\href@noop {} {\bibfield  {journal} {\bibinfo  {journal} {Phys.
  Rev.}\ }\textbf {\bibinfo {volume} {69}},\ \bibinfo {pages} {681} (\bibinfo
  {year} {1946})}\BibitemShut {NoStop}%
\bibitem [{\citenamefont {Novotny}\ and\ \citenamefont
  {Hecht}(2006)}]{Novotny2006}%
  \BibitemOpen
  \bibfield  {author} {\bibinfo {author} {\bibfnamefont {L.}~\bibnamefont
  {Novotny}}\ and\ \bibinfo {author} {\bibfnamefont {B.}~\bibnamefont
  {Hecht}},\ }\href@noop {} {\emph {\bibinfo {title} {Principles of
  Nano-Optics}}}\ (\bibinfo  {publisher} {Cambridge University Press,
  Cambridge},\ \bibinfo {year} {2006})\BibitemShut {NoStop}%
\bibitem [{\citenamefont {Drexhage}(1970)}]{Drexhage1970}%
  \BibitemOpen
  \bibfield  {author} {\bibinfo {author} {\bibfnamefont {K.~H.}\ \bibnamefont
  {Drexhage}},\ }\href@noop {} {\bibfield  {journal} {\bibinfo  {journal} {J.
  of Lumin.}\ }\textbf {\bibinfo {volume} {1-2}},\ \bibinfo {pages} {693}
  (\bibinfo {year} {1970})}\BibitemShut {NoStop}%
\bibitem [{\citenamefont {Snoeks}\ \emph {et~al.}(1995)\citenamefont {Snoeks},
  \citenamefont {Lagendijk},\ and\ \citenamefont {Polman}}]{Snoeks1995}%
  \BibitemOpen
  \bibfield  {author} {\bibinfo {author} {\bibfnamefont {E.}~\bibnamefont
  {Snoeks}}, \bibinfo {author} {\bibfnamefont {A.}~\bibnamefont {Lagendijk}}, \
  and\ \bibinfo {author} {\bibfnamefont {A.}~\bibnamefont {Polman}},\ }\href
  {\doibase 10.1103/PhysRevLett.74.2459} {\bibfield  {journal} {\bibinfo
  {journal} {Phys. Rev. Lett.}\ }\textbf {\bibinfo {volume} {74}},\ \bibinfo
  {pages} {2459} (\bibinfo {year} {1995})}\BibitemShut {NoStop}%
\bibitem [{\citenamefont {Lodahl}\ \emph {et~al.}(2004)\citenamefont {Lodahl},
  \citenamefont {van Driel}, \citenamefont {Nikolaev}, \citenamefont {Irman},
  \citenamefont {Overgaag}, \citenamefont {Vanmaekelbergh},\ and\ \citenamefont
  {Vos}}]{Lodahl2004}%
  \BibitemOpen
  \bibfield  {author} {\bibinfo {author} {\bibfnamefont {P.}~\bibnamefont
  {Lodahl}}, \bibinfo {author} {\bibfnamefont {A.~F.}\ \bibnamefont {van
  Driel}}, \bibinfo {author} {\bibfnamefont {I.~S.}\ \bibnamefont {Nikolaev}},
  \bibinfo {author} {\bibfnamefont {A.}~\bibnamefont {Irman}}, \bibinfo
  {author} {\bibfnamefont {K.}~\bibnamefont {Overgaag}}, \bibinfo {author}
  {\bibfnamefont {D.}~\bibnamefont {Vanmaekelbergh}}, \ and\ \bibinfo {author}
  {\bibfnamefont {W.~L.}\ \bibnamefont {Vos}},\ }\href@noop {} {\bibfield
  {journal} {\bibinfo  {journal} {Nature}\ }\textbf {\bibinfo {volume} {430}},\
  \bibinfo {pages} {654} (\bibinfo {year} {2004})}\BibitemShut {NoStop}%
\bibitem [{\citenamefont {Barth}\ \emph {et~al.}(2009)\citenamefont {Barth},
  \citenamefont {N\"{u}sse}, \citenamefont {L\"{o}chel},\ and\ \citenamefont
  {Benson}}]{Barth2009}%
  \BibitemOpen
  \bibfield  {author} {\bibinfo {author} {\bibfnamefont {M.}~\bibnamefont
  {Barth}}, \bibinfo {author} {\bibfnamefont {N.}~\bibnamefont {N\"{u}sse}},
  \bibinfo {author} {\bibfnamefont {B.}~\bibnamefont {L\"{o}chel}}, \ and\
  \bibinfo {author} {\bibfnamefont {O.}~\bibnamefont {Benson}},\ }\href@noop {}
  {\bibfield  {journal} {\bibinfo  {journal} {Opt. Lett.}\ }\textbf {\bibinfo
  {volume} {34}},\ \bibinfo {pages} {1108} (\bibinfo {year}
  {2009})}\BibitemShut {NoStop}%
\bibitem [{\citenamefont {van~der Sar}\ \emph {et~al.}(2011)\citenamefont
  {van~der Sar}, \citenamefont {Hagemeier}, \citenamefont {Pfaff},
  \citenamefont {Heeres}, \citenamefont {Thon}, \citenamefont {Kim},
  \citenamefont {Petroff}, \citenamefont {Oosterkamp}, \citenamefont
  {Bouwmeester},\ and\ \citenamefont {Hanson}}]{Sar2011}%
  \BibitemOpen
  \bibfield  {author} {\bibinfo {author} {\bibfnamefont {T.}~\bibnamefont
  {van~der Sar}}, \bibinfo {author} {\bibfnamefont {J.}~\bibnamefont
  {Hagemeier}}, \bibinfo {author} {\bibfnamefont {W.}~\bibnamefont {Pfaff}},
  \bibinfo {author} {\bibfnamefont {E.~C.}\ \bibnamefont {Heeres}}, \bibinfo
  {author} {\bibfnamefont {S.~M.}\ \bibnamefont {Thon}}, \bibinfo {author}
  {\bibfnamefont {H.}~\bibnamefont {Kim}}, \bibinfo {author} {\bibfnamefont
  {P.~M.}\ \bibnamefont {Petroff}}, \bibinfo {author} {\bibfnamefont {T.~H.}\
  \bibnamefont {Oosterkamp}}, \bibinfo {author} {\bibfnamefont
  {D.}~\bibnamefont {Bouwmeester}}, \ and\ \bibinfo {author} {\bibfnamefont
  {R.}~\bibnamefont {Hanson}},\ }\href {\doibase 10.1063/1.3571437} {\bibfield
  {journal} {\bibinfo  {journal} {Appl. Phys. Lett.}\ }\textbf {\bibinfo
  {volume} {98}},\ \bibinfo {eid} {193103} (\bibinfo {year}
  {2011})}\BibitemShut {NoStop}%
\bibitem [{\citenamefont {Vahala}(2005)}]{Vahala2005}%
  \BibitemOpen
  \bibinfo {editor} {\bibfnamefont {K.}~\bibnamefont {Vahala}},\ ed.,\
  \href@noop {} {\emph {\bibinfo {title} {Optical Microcavities (Advanced
  Series in Applied Physics, Vol. 5)}}}\ (\bibinfo  {publisher} {World
  Scientific, Singapore},\ \bibinfo {year} {2005})\BibitemShut {NoStop}%
\bibitem [{\citenamefont {Anger}\ \emph {et~al.}(2006)\citenamefont {Anger},
  \citenamefont {Bharadwaj},\ and\ \citenamefont {Novotny}}]{Anger2006}%
  \BibitemOpen
  \bibfield  {author} {\bibinfo {author} {\bibfnamefont {P.}~\bibnamefont
  {Anger}}, \bibinfo {author} {\bibfnamefont {P.}~\bibnamefont {Bharadwaj}}, \
  and\ \bibinfo {author} {\bibfnamefont {L.}~\bibnamefont {Novotny}},\
  }\href@noop {} {\bibfield  {journal} {\bibinfo  {journal} {Phys. Rev. Lett.}\
  }\textbf {\bibinfo {volume} {96}},\ \bibinfo {pages} {113002} (\bibinfo
  {year} {2006})}\BibitemShut {NoStop}%
\bibitem [{\citenamefont {K\"{u}hn}\ \emph {et~al.}(2006)\citenamefont
  {K\"{u}hn}, \citenamefont {H\r{a}kanson}, \citenamefont {Rogobete},\ and\
  \citenamefont {Sandoghdar}}]{Kuehn2006}%
  \BibitemOpen
  \bibfield  {author} {\bibinfo {author} {\bibfnamefont {S.}~\bibnamefont
  {K\"{u}hn}}, \bibinfo {author} {\bibfnamefont {U.}~\bibnamefont
  {H\r{a}kanson}}, \bibinfo {author} {\bibfnamefont {L.}~\bibnamefont
  {Rogobete}}, \ and\ \bibinfo {author} {\bibfnamefont {V.}~\bibnamefont
  {Sandoghdar}},\ }\href {\doibase 10.1103/PhysRevLett.97.017402} {\bibfield
  {journal} {\bibinfo  {journal} {Phys. Rev. Lett.}\ }\textbf {\bibinfo
  {volume} {97}},\ \bibinfo {pages} {017402} (\bibinfo {year}
  {2006})}\BibitemShut {NoStop}%
\bibitem [{\citenamefont {Schietinger}\ \emph {et~al.}(2009)\citenamefont
  {Schietinger}, \citenamefont {Barth}, \citenamefont {Aichele},\ and\
  \citenamefont {Benson}}]{Schietinger2009}%
  \BibitemOpen
  \bibfield  {author} {\bibinfo {author} {\bibfnamefont {S.}~\bibnamefont
  {Schietinger}}, \bibinfo {author} {\bibfnamefont {M.}~\bibnamefont {Barth}},
  \bibinfo {author} {\bibfnamefont {T.}~\bibnamefont {Aichele}}, \ and\
  \bibinfo {author} {\bibfnamefont {O.}~\bibnamefont {Benson}},\ }\href@noop {}
  {\bibfield  {journal} {\bibinfo  {journal} {Nano Lett.}\ }\textbf {\bibinfo
  {volume} {9}},\ \bibinfo {pages} {1694} (\bibinfo {year} {2009})}\BibitemShut
  {NoStop}%
\bibitem [{\citenamefont {Kinkhabwala}\ \emph {et~al.}(2009)\citenamefont
  {Kinkhabwala}, \citenamefont {Yu}, \citenamefont {Fan}, \citenamefont
  {Avlasevich}, \citenamefont {M\"{u}llen},\ and\ \citenamefont
  {Moerner}}]{Kinkhabwala2009}%
  \BibitemOpen
  \bibfield  {author} {\bibinfo {author} {\bibfnamefont {A.}~\bibnamefont
  {Kinkhabwala}}, \bibinfo {author} {\bibfnamefont {Z.}~\bibnamefont {Yu}},
  \bibinfo {author} {\bibfnamefont {S.}~\bibnamefont {Fan}}, \bibinfo {author}
  {\bibfnamefont {Y.}~\bibnamefont {Avlasevich}}, \bibinfo {author}
  {\bibfnamefont {K.}~\bibnamefont {M\"{u}llen}}, \ and\ \bibinfo {author}
  {\bibfnamefont {W.~E.}\ \bibnamefont {Moerner}},\ }\href@noop {} {\bibfield
  {journal} {\bibinfo  {journal} {Nature Photon.}\ }\textbf {\bibinfo {volume}
  {3}},\ \bibinfo {pages} {654} (\bibinfo {year} {2009})}\BibitemShut {NoStop}%
\bibitem [{\citenamefont {Farahani}\ \emph {et~al.}(2005)\citenamefont
  {Farahani}, \citenamefont {Pohl}, \citenamefont {Eisler},\ and\ \citenamefont
  {Hecht}}]{Farahani2005}%
  \BibitemOpen
  \bibfield  {author} {\bibinfo {author} {\bibfnamefont {J.~N.}\ \bibnamefont
  {Farahani}}, \bibinfo {author} {\bibfnamefont {D.~W.}\ \bibnamefont {Pohl}},
  \bibinfo {author} {\bibfnamefont {H.-J.}\ \bibnamefont {Eisler}}, \ and\
  \bibinfo {author} {\bibfnamefont {B.}~\bibnamefont {Hecht}},\ }\href@noop {}
  {\bibfield  {journal} {\bibinfo  {journal} {Phys. Rev. Lett.}\ }\textbf
  {\bibinfo {volume} {95}},\ \bibinfo {pages} {017402} (\bibinfo {year}
  {2005})}\BibitemShut {NoStop}%
\bibitem [{\citenamefont {Benson}(2011)}]{Benson2011}%
  \BibitemOpen
  \bibfield  {author} {\bibinfo {author} {\bibfnamefont {O.}~\bibnamefont
  {Benson}},\ }\href@noop {} {\bibfield  {journal} {\bibinfo  {journal}
  {Nature}\ }\textbf {\bibinfo {volume} {480}},\ \bibinfo {pages} {193}
  (\bibinfo {year} {2011})}\BibitemShut {NoStop}%
\bibitem [{\citenamefont {Xiao}\ \emph {et~al.}(2012)\citenamefont {Xiao},
  \citenamefont {Liu}, \citenamefont {Li}, \citenamefont {Chen}, \citenamefont
  {Li},\ and\ \citenamefont {Gong}}]{Xiao2012}%
  \BibitemOpen
  \bibfield  {author} {\bibinfo {author} {\bibfnamefont {Y.-F.}\ \bibnamefont
  {Xiao}}, \bibinfo {author} {\bibfnamefont {Y.-C.}\ \bibnamefont {Liu}},
  \bibinfo {author} {\bibfnamefont {B.-B.}\ \bibnamefont {Li}}, \bibinfo
  {author} {\bibfnamefont {Y.-L.}\ \bibnamefont {Chen}}, \bibinfo {author}
  {\bibfnamefont {Y.}~\bibnamefont {Li}}, \ and\ \bibinfo {author}
  {\bibfnamefont {Q.}~\bibnamefont {Gong}},\ }\href {\doibase
  10.1103/PhysRevA.85.031805} {\bibfield  {journal} {\bibinfo  {journal} {Phys.
  Rev. A}\ }\textbf {\bibinfo {volume} {85}},\ \bibinfo {pages} {031805}
  (\bibinfo {year} {2012})}\BibitemShut {NoStop}%
\bibitem [{\citenamefont {Engheta}\ \emph {et~al.}(2005)\citenamefont
  {Engheta}, \citenamefont {Salandrino},\ and\ \citenamefont
  {Al\`{u}}}]{Engheta2005}%
  \BibitemOpen
  \bibfield  {author} {\bibinfo {author} {\bibfnamefont {N.}~\bibnamefont
  {Engheta}}, \bibinfo {author} {\bibfnamefont {A.}~\bibnamefont {Salandrino}},
  \ and\ \bibinfo {author} {\bibfnamefont {A.}~\bibnamefont {Al\`{u}}},\
  }\href@noop {} {\bibfield  {journal} {\bibinfo  {journal} {Phys. Rev. Lett.}\
  }\textbf {\bibinfo {volume} {95}},\ \bibinfo {pages} {095504} (\bibinfo
  {year} {2005})}\BibitemShut {NoStop}%
\bibitem [{\citenamefont {Engheta}(2007)}]{Engheta2007}%
  \BibitemOpen
  \bibfield  {author} {\bibinfo {author} {\bibfnamefont {N.}~\bibnamefont
  {Engheta}},\ }\href {\doibase 10.1126/science.1133268} {\bibfield  {journal}
  {\bibinfo  {journal} {Science}\ }\textbf {\bibinfo {volume} {317}},\ \bibinfo
  {pages} {1698} (\bibinfo {year} {2007})}\BibitemShut {NoStop}%
\bibitem [{\citenamefont {Al\`{u}}\ and\ \citenamefont
  {Engheta}(2008{\natexlab{a}})}]{Al`u2008}%
  \BibitemOpen
  \bibfield  {author} {\bibinfo {author} {\bibfnamefont {A.}~\bibnamefont
  {Al\`{u}}}\ and\ \bibinfo {author} {\bibfnamefont {N.}~\bibnamefont
  {Engheta}},\ }\href {\doibase 10.1103/PhysRevB.78.195111} {\bibfield
  {journal} {\bibinfo  {journal} {Phys. Rev. B}\ }\textbf {\bibinfo {volume}
  {78}},\ \bibinfo {pages} {195111} (\bibinfo {year}
  {2008}{\natexlab{a}})}\BibitemShut {NoStop}%
\bibitem [{\citenamefont {Al\`{u}}\ and\ \citenamefont
  {Engheta}(2008{\natexlab{b}})}]{Al`u2008b}%
  \BibitemOpen
  \bibfield  {author} {\bibinfo {author} {\bibfnamefont {A.}~\bibnamefont
  {Al\`{u}}}\ and\ \bibinfo {author} {\bibfnamefont {N.}~\bibnamefont
  {Engheta}},\ }\href {\doibase 10.1103/PhysRevLett.101.043901} {\bibfield
  {journal} {\bibinfo  {journal} {Phys. Rev. Lett.}\ }\textbf {\bibinfo
  {volume} {101}},\ \bibinfo {pages} {043901} (\bibinfo {year}
  {2008}{\natexlab{b}})}\BibitemShut {NoStop}%
\bibitem [{\citenamefont {Greffet}\ \emph {et~al.}(2010)\citenamefont
  {Greffet}, \citenamefont {Laroche},\ and\ \citenamefont
  {Marquier}}]{Greffet2010}%
  \BibitemOpen
  \bibfield  {author} {\bibinfo {author} {\bibfnamefont {J.-J.}\ \bibnamefont
  {Greffet}}, \bibinfo {author} {\bibfnamefont {M.}~\bibnamefont {Laroche}}, \
  and\ \bibinfo {author} {\bibfnamefont {F.}~\bibnamefont {Marquier}},\ }\href
  {\doibase 10.1103/PhysRevLett.105.117701} {\bibfield  {journal} {\bibinfo
  {journal} {Phys. Rev. Lett.}\ }\textbf {\bibinfo {volume} {105}},\ \bibinfo
  {pages} {117701} (\bibinfo {year} {2010})}\BibitemShut {NoStop}%
\bibitem [{Note1()}]{Note1}%
  \BibitemOpen
  \bibinfo {note} {We assume the source to be aligned radially to an isotropic
  particle. For a full vectorial expression see Ref.~\protect \rev@citealp
  {Sipe1974}.}\BibitemShut {Stop}%
\bibitem [{\citenamefont {Mertens}\ \emph {et~al.}(2007)\citenamefont
  {Mertens}, \citenamefont {Koenderink},\ and\ \citenamefont
  {Polman}}]{Mertens2007}%
  \BibitemOpen
  \bibfield  {author} {\bibinfo {author} {\bibfnamefont {H.}~\bibnamefont
  {Mertens}}, \bibinfo {author} {\bibfnamefont {A.~F.}\ \bibnamefont
  {Koenderink}}, \ and\ \bibinfo {author} {\bibfnamefont {A.}~\bibnamefont
  {Polman}},\ }\href {\doibase 10.1103/PhysRevB.76.115123} {\bibfield
  {journal} {\bibinfo  {journal} {Phys. Rev. B}\ }\textbf {\bibinfo {volume}
  {76}},\ \bibinfo {pages} {115123} (\bibinfo {year} {2007})}\BibitemShut
  {NoStop}%
\bibitem [{\citenamefont {Weber}\ and\ \citenamefont {Ford}(2004)}]{Weber2004}%
  \BibitemOpen
  \bibfield  {author} {\bibinfo {author} {\bibfnamefont {W.~H.}\ \bibnamefont
  {Weber}}\ and\ \bibinfo {author} {\bibfnamefont {G.~W.}\ \bibnamefont
  {Ford}},\ }\href@noop {} {\bibfield  {journal} {\bibinfo  {journal} {Physical
  Review B}\ }\textbf {\bibinfo {volume} {70}},\ \bibinfo {pages} {125429}
  (\bibinfo {year} {2004})}\BibitemShut {NoStop}%
\bibitem [{\citenamefont {Garc\'{i}a~de Abajo}(2007)}]{GarciadeAbajo2007}%
  \BibitemOpen
  \bibfield  {author} {\bibinfo {author} {\bibfnamefont {F.~J.}\ \bibnamefont
  {Garc\'{i}a~de Abajo}},\ }\href {\doibase 10.1103/RevModPhys.79.1267}
  {\bibfield  {journal} {\bibinfo  {journal} {Rev. Mod. Phys.}\ }\textbf
  {\bibinfo {volume} {79}},\ \bibinfo {pages} {1267} (\bibinfo {year}
  {2007})}\BibitemShut {NoStop}%
\bibitem [{Note2()}]{Note2}%
  \BibitemOpen
  \bibinfo {note} {Since $\protect \text {Re}\protect \boldsymbol {G}_{\protect
  \text {vac}}$ diverges at the source, and one is only interested in relative
  frequency shifts, it is commonly included in $\protect \boldsymbol {\alpha
  }_0$ to yield a finite resonance frequency $\omega _0$.}\BibitemShut {Stop}%
\bibitem [{\citenamefont {Tai}(1994)}]{Tai1994}%
  \BibitemOpen
  \bibfield  {author} {\bibinfo {author} {\bibfnamefont {C.-T.}\ \bibnamefont
  {Tai}},\ }\href@noop {} {\emph {\bibinfo {title} {Dyadic Green Functions in
  Electromagnetic Theory}}}\ (\bibinfo  {publisher} {IEEE Press},\ \bibinfo
  {year} {1994})\BibitemShut {NoStop}%
\bibitem [{\citenamefont {Paulus}\ \emph {et~al.}(2000)\citenamefont {Paulus},
  \citenamefont {Gay-Balmaz},\ and\ \citenamefont {Martin}}]{Paulus2000}%
  \BibitemOpen
  \bibfield  {author} {\bibinfo {author} {\bibfnamefont {M.}~\bibnamefont
  {Paulus}}, \bibinfo {author} {\bibfnamefont {P.}~\bibnamefont {Gay-Balmaz}},
  \ and\ \bibinfo {author} {\bibfnamefont {O.~J.~F.}\ \bibnamefont {Martin}},\
  }\href {\doibase 10.1103/PhysRevE.62.5797} {\bibfield  {journal} {\bibinfo
  {journal} {Phys. Rev. E}\ }\textbf {\bibinfo {volume} {62}},\ \bibinfo
  {pages} {5797} (\bibinfo {year} {2000})}\BibitemShut {NoStop}%
\bibitem [{Note3()}]{Note3}%
  \BibitemOpen
  \bibinfo {note} {For spherical particles of volume $V$ and dielectric
  constant $\epsilon (\omega )$, the electrostatic polarizability is simply
  $3V(\epsilon -1)/(\epsilon +2)$. A Drude model for $\epsilon $ yields a
  Lorentzian. We choose $\omega _0$=$4.76\times 10^{15}\protect \tmspace
  +\thinmuskip {.1667em}\protect \text {s}^{-1}$ and $\gamma $=$8.3\times
  10^{12}\protect \tmspace +\thinmuskip {.1667em}\protect \text {s}^{-1}$ to
  model the Ag particles.}\BibitemShut {Stop}%
\bibitem [{\citenamefont {Koenderink}(2010)}]{Koenderink2010}%
  \BibitemOpen
  \bibfield  {author} {\bibinfo {author} {\bibfnamefont {A.~F.}\ \bibnamefont
  {Koenderink}},\ }\href {\doibase 10.1364/OL.35.004208} {\bibfield  {journal}
  {\bibinfo  {journal} {Opt. Lett.}\ }\textbf {\bibinfo {volume} {35}},\
  \bibinfo {pages} {4208} (\bibinfo {year} {2010})}\BibitemShut {NoStop}%
\bibitem [{\citenamefont {Prodan}\ \emph {et~al.}(2003)\citenamefont {Prodan},
  \citenamefont {Radloff}, \citenamefont {Halas},\ and\ \citenamefont
  {Nordlander}}]{Prodan2003}%
  \BibitemOpen
  \bibfield  {author} {\bibinfo {author} {\bibfnamefont {E.}~\bibnamefont
  {Prodan}}, \bibinfo {author} {\bibfnamefont {C.}~\bibnamefont {Radloff}},
  \bibinfo {author} {\bibfnamefont {N.~J.}\ \bibnamefont {Halas}}, \ and\
  \bibinfo {author} {\bibfnamefont {P.}~\bibnamefont {Nordlander}},\
  }\href@noop {} {\bibfield  {journal} {\bibinfo  {journal} {Science}\ }\textbf
  {\bibinfo {volume} {302}},\ \bibinfo {pages} {419} (\bibinfo {year}
  {2003})}\BibitemShut {NoStop}%
\bibitem [{\citenamefont {Bohren}\ and\ \citenamefont
  {Huffman}(1983)}]{Bohren1983}%
  \BibitemOpen
  \bibfield  {author} {\bibinfo {author} {\bibfnamefont {C.~F.}\ \bibnamefont
  {Bohren}}\ and\ \bibinfo {author} {\bibfnamefont {D.~R.}\ \bibnamefont
  {Huffman}},\ }\href@noop {} {\emph {\bibinfo {title} {Absorption and
  Scattering of Light by Small Particles}}}\ (\bibinfo  {publisher} {John Wiley
  and Sons, New York},\ \bibinfo {year} {1983})\BibitemShut {NoStop}%
\bibitem [{\citenamefont {Agha}\ \emph {et~al.}(2006)\citenamefont {Agha},
  \citenamefont {Sharping}, \citenamefont {Foster},\ and\ \citenamefont
  {Gaeta}}]{Agha2006}%
  \BibitemOpen
  \bibfield  {author} {\bibinfo {author} {\bibfnamefont {I.}~\bibnamefont
  {Agha}}, \bibinfo {author} {\bibfnamefont {J.}~\bibnamefont {Sharping}},
  \bibinfo {author} {\bibfnamefont {M.}~\bibnamefont {Foster}}, \ and\ \bibinfo
  {author} {\bibfnamefont {A.}~\bibnamefont {Gaeta}},\ }\href@noop {}
  {\bibfield  {journal} {\bibinfo  {journal} {Appl. Phys. B}\ }\textbf
  {\bibinfo {volume} {83}},\ \bibinfo {pages} {303} (\bibinfo {year}
  {2006})}\BibitemShut {NoStop}%
\bibitem [{\citenamefont {Koenderink}\ \emph {et~al.}(2005)\citenamefont
  {Koenderink}, \citenamefont {Kafesaki}, \citenamefont {Buchler},\ and\
  \citenamefont {Sandoghdar}}]{Koenderink2005a}%
  \BibitemOpen
  \bibfield  {author} {\bibinfo {author} {\bibfnamefont {A.~F.}\ \bibnamefont
  {Koenderink}}, \bibinfo {author} {\bibfnamefont {M.}~\bibnamefont
  {Kafesaki}}, \bibinfo {author} {\bibfnamefont {B.~C.}\ \bibnamefont
  {Buchler}}, \ and\ \bibinfo {author} {\bibfnamefont {V.}~\bibnamefont
  {Sandoghdar}},\ }\href {\doibase 10.1103/PhysRevLett.95.153904} {\bibfield
  {journal} {\bibinfo  {journal} {Phys. Rev. Lett.}\ }\textbf {\bibinfo
  {volume} {95}},\ \bibinfo {pages} {153904} (\bibinfo {year}
  {2005})}\BibitemShut {NoStop}%
\bibitem [{Note4()}]{Note4}%
  \BibitemOpen
  \bibinfo {note} {It is easy to show that a small dimer antenna is described
  as a single scatterer with $\protect \boldsymbol {\alpha }_\protect \text
  {eff}$=$2\left [\protect \boldsymbol {\alpha }_\protect \text
  {B}^{-1}-\protect \boldsymbol {G}_\protect \text {B}(\protect \boldsymbol
  {r}_1,\protect \boldsymbol {r}_2)\right ]^{-1}$}\BibitemShut {NoStop}%
\bibitem [{Note5()}]{Note5}%
  \BibitemOpen
  \bibinfo {note} {For a full vectorial expression of the optical theorem see
  Ref.~\protect \rev@citealp {Sipe1974}.}\BibitemShut {Stop}%
\bibitem [{\citenamefont {Newton}(2002)}]{Newton2002}%
  \BibitemOpen
  \bibfield  {author} {\bibinfo {author} {\bibfnamefont {R.~G.}\ \bibnamefont
  {Newton}},\ }\href@noop {} {\emph {\bibinfo {title} {Scattering Theory of
  Waves and Particles}}}\ (\bibinfo  {publisher} {Dover Publications, New
  York},\ \bibinfo {year} {2002})\BibitemShut {NoStop}%
\bibitem [{\citenamefont {Buchler}\ \emph {et~al.}(2005)\citenamefont
  {Buchler}, \citenamefont {Kalkbrenner}, \citenamefont {Hettich},\ and\
  \citenamefont {Sandoghdar}}]{Buchler2005}%
  \BibitemOpen
  \bibfield  {author} {\bibinfo {author} {\bibfnamefont {B.~C.}\ \bibnamefont
  {Buchler}}, \bibinfo {author} {\bibfnamefont {T.}~\bibnamefont
  {Kalkbrenner}}, \bibinfo {author} {\bibfnamefont {C.}~\bibnamefont
  {Hettich}}, \ and\ \bibinfo {author} {\bibfnamefont {V.}~\bibnamefont
  {Sandoghdar}},\ }\href {\doibase 10.1103/PhysRevLett.95.063003} {\bibfield
  {journal} {\bibinfo  {journal} {Phys. Rev. Lett.}\ }\textbf {\bibinfo
  {volume} {95}},\ \bibinfo {pages} {063003} (\bibinfo {year}
  {2005})}\BibitemShut {NoStop}%
\bibitem [{Note6()}]{Note6}%
  \BibitemOpen
  \bibinfo {note} {Note that $\protect \boldsymbol {G}_{\protect \text
  {vac}}\propto 1/d^3$ while $\protect \boldsymbol {G}_{\protect \text {B}}$ is
  of order $k^3$.}\BibitemShut {Stop}%
\bibitem [{\citenamefont {Frimmer}\ \emph {et~al.}(2012)\citenamefont
  {Frimmer}, \citenamefont {Coenen},\ and\ \citenamefont
  {Koenderink}}]{Frimmer2012}%
  \BibitemOpen
  \bibfield  {author} {\bibinfo {author} {\bibfnamefont {M.}~\bibnamefont
  {Frimmer}}, \bibinfo {author} {\bibfnamefont {T.}~\bibnamefont {Coenen}}, \
  and\ \bibinfo {author} {\bibfnamefont {A.~F.}\ \bibnamefont {Koenderink}},\
  }\href {\doibase 10.1103/PhysRevLett.108.077404} {\bibfield  {journal}
  {\bibinfo  {journal} {Phys. Rev. Lett.}\ }\textbf {\bibinfo {volume} {108}},\
  \bibinfo {pages} {077404} (\bibinfo {year} {2012})}\BibitemShut {NoStop}%
\bibitem [{\citenamefont {Motsch}\ \emph {et~al.}(2010)\citenamefont {Motsch},
  \citenamefont {Zeppenfeld}, \citenamefont {Pinkse},\ and\ \citenamefont
  {Rempe}}]{Motsch2010}%
  \BibitemOpen
  \bibfield  {author} {\bibinfo {author} {\bibfnamefont {M.}~\bibnamefont
  {Motsch}}, \bibinfo {author} {\bibfnamefont {M.}~\bibnamefont {Zeppenfeld}},
  \bibinfo {author} {\bibfnamefont {P.~W.~H.}\ \bibnamefont {Pinkse}}, \ and\
  \bibinfo {author} {\bibfnamefont {G.}~\bibnamefont {Rempe}},\ }\href@noop {}
  {\bibfield  {journal} {\bibinfo  {journal} {New J. Phys.}\ }\textbf {\bibinfo
  {volume} {12}},\ \bibinfo {pages} {063022} (\bibinfo {year}
  {2010})}\BibitemShut {NoStop}%
\bibitem [{\citenamefont {Kalkbrenner}\ \emph {et~al.}(2005)\citenamefont
  {Kalkbrenner}, \citenamefont {H\aa{}kanson}, \citenamefont {Sch\"adle},
  \citenamefont {Burger}, \citenamefont {Henkel},\ and\ \citenamefont
  {Sandoghdar}}]{Kalkbrenner2005}%
  \BibitemOpen
  \bibfield  {author} {\bibinfo {author} {\bibfnamefont {T.}~\bibnamefont
  {Kalkbrenner}}, \bibinfo {author} {\bibfnamefont {U.}~\bibnamefont
  {H\aa{}kanson}}, \bibinfo {author} {\bibfnamefont {A.}~\bibnamefont
  {Sch\"adle}}, \bibinfo {author} {\bibfnamefont {S.}~\bibnamefont {Burger}},
  \bibinfo {author} {\bibfnamefont {C.}~\bibnamefont {Henkel}}, \ and\ \bibinfo
  {author} {\bibfnamefont {V.}~\bibnamefont {Sandoghdar}},\ }\href {\doibase
  10.1103/PhysRevLett.95.200801} {\bibfield  {journal} {\bibinfo  {journal}
  {Phys. Rev. Lett.}\ }\textbf {\bibinfo {volume} {95}},\ \bibinfo {pages}
  {200801} (\bibinfo {year} {2005})}\BibitemShut {NoStop}%
\bibitem [{\citenamefont {Castani\'{e}}\ \emph {et~al.}(2012)\citenamefont
  {Castani\'{e}}, \citenamefont {Vincent}, \citenamefont {Pierrat},\ and\
  \citenamefont {Carminati}}]{Castanie2011}%
  \BibitemOpen
  \bibfield  {author} {\bibinfo {author} {\bibfnamefont {E.}~\bibnamefont
  {Castani\'{e}}}, \bibinfo {author} {\bibfnamefont {R.}~\bibnamefont
  {Vincent}}, \bibinfo {author} {\bibfnamefont {R.}~\bibnamefont {Pierrat}}, \
  and\ \bibinfo {author} {\bibfnamefont {R.}~\bibnamefont {Carminati}},\
  }\href@noop {} {\bibfield  {journal} {\bibinfo  {journal} {Intern. J. of
  Opt.}\ }\textbf {\bibinfo {volume} {2012}},\ \bibinfo {pages} {452047}
  (\bibinfo {year} {2012})}\BibitemShut {NoStop}%
\bibitem [{\citenamefont {Andersen}\ \emph {et~al.}(2011)\citenamefont
  {Andersen}, \citenamefont {Stobbe}, \citenamefont {S{\o}rensen},\ and\
  \citenamefont {Lodahl}}]{Andersen2011}%
  \BibitemOpen
  \bibfield  {author} {\bibinfo {author} {\bibfnamefont {M.~L.}\ \bibnamefont
  {Andersen}}, \bibinfo {author} {\bibfnamefont {S.}~\bibnamefont {Stobbe}},
  \bibinfo {author} {\bibfnamefont {A.~S.}\ \bibnamefont {S{\o}rensen}}, \ and\
  \bibinfo {author} {\bibfnamefont {P.}~\bibnamefont {Lodahl}},\ }\href@noop {}
  {\bibfield  {journal} {\bibinfo  {journal} {Nature Phys.}\ }\textbf {\bibinfo
  {volume} {7}},\ \bibinfo {pages} {215} (\bibinfo {year} {2011})}\BibitemShut
  {NoStop}%
\bibitem [{\citenamefont {Verslegers}\ \emph {et~al.}(2012)\citenamefont
  {Verslegers}, \citenamefont {Yu}, \citenamefont {Ruan}, \citenamefont
  {Catrysse},\ and\ \citenamefont {Fan}}]{Verslegers2012}%
  \BibitemOpen
  \bibfield  {author} {\bibinfo {author} {\bibfnamefont {L.}~\bibnamefont
  {Verslegers}}, \bibinfo {author} {\bibfnamefont {Z.}~\bibnamefont {Yu}},
  \bibinfo {author} {\bibfnamefont {Z.}~\bibnamefont {Ruan}}, \bibinfo {author}
  {\bibfnamefont {P.~B.}\ \bibnamefont {Catrysse}}, \ and\ \bibinfo {author}
  {\bibfnamefont {S.}~\bibnamefont {Fan}},\ }\href {\doibase
  10.1103/PhysRevLett.108.083902} {\bibfield  {journal} {\bibinfo  {journal}
  {Phys. Rev. Lett.}\ }\textbf {\bibinfo {volume} {108}},\ \bibinfo {pages}
  {083902} (\bibinfo {year} {2012})}\BibitemShut {NoStop}%
\bibitem [{\citenamefont {Mazzei}\ \emph {et~al.}(2007)\citenamefont {Mazzei},
  \citenamefont {G\"otzinger}, \citenamefont {de~S.~Menezes}, \citenamefont
  {Zumofen}, \citenamefont {Benson},\ and\ \citenamefont
  {Sandoghdar}}]{Mazzei2007}%
  \BibitemOpen
  \bibfield  {author} {\bibinfo {author} {\bibfnamefont {A.}~\bibnamefont
  {Mazzei}}, \bibinfo {author} {\bibfnamefont {S.}~\bibnamefont {G\"otzinger}},
  \bibinfo {author} {\bibfnamefont {L.}~\bibnamefont {de~S.~Menezes}}, \bibinfo
  {author} {\bibfnamefont {G.}~\bibnamefont {Zumofen}}, \bibinfo {author}
  {\bibfnamefont {O.}~\bibnamefont {Benson}}, \ and\ \bibinfo {author}
  {\bibfnamefont {V.}~\bibnamefont {Sandoghdar}},\ }\href {\doibase
  10.1103/PhysRevLett.99.173603} {\bibfield  {journal} {\bibinfo  {journal}
  {Phys. Rev. Lett.}\ }\textbf {\bibinfo {volume} {99}},\ \bibinfo {pages}
  {173603} (\bibinfo {year} {2007})}\BibitemShut {NoStop}%
\bibitem [{\citenamefont {Wilde}\ \emph {et~al.}(2006)\citenamefont {Wilde},
  \citenamefont {Formanek}, \citenamefont {Carminati}, \citenamefont {Gralak},
  \citenamefont {Lemoine}, \citenamefont {Joulain}, \citenamefont {Mule},
  \citenamefont {Chen},\ and\ \citenamefont {Greffet}}]{Wilde2006}%
  \BibitemOpen
  \bibfield  {author} {\bibinfo {author} {\bibfnamefont {Y.~D.}\ \bibnamefont
  {Wilde}}, \bibinfo {author} {\bibfnamefont {F.}~\bibnamefont {Formanek}},
  \bibinfo {author} {\bibfnamefont {R.}~\bibnamefont {Carminati}}, \bibinfo
  {author} {\bibfnamefont {B.}~\bibnamefont {Gralak}}, \bibinfo {author}
  {\bibfnamefont {P.-A.}\ \bibnamefont {Lemoine}}, \bibinfo {author}
  {\bibfnamefont {K.}~\bibnamefont {Joulain}}, \bibinfo {author} {\bibfnamefont
  {J.-P.}\ \bibnamefont {Mule}}, \bibinfo {author} {\bibfnamefont
  {Y.}~\bibnamefont {Chen}}, \ and\ \bibinfo {author} {\bibfnamefont {J.-J.}\
  \bibnamefont {Greffet}},\ }\href@noop {} {\bibfield  {journal} {\bibinfo
  {journal} {Nature}\ }\textbf {\bibinfo {volume} {444}},\ \bibinfo {pages}
  {740} (\bibinfo {year} {2006})}\BibitemShut {NoStop}%
\bibitem [{\citenamefont {Liu}\ \emph {et~al.}(2011)\citenamefont {Liu},
  \citenamefont {Tyler}, \citenamefont {Starr}, \citenamefont {Starr},
  \citenamefont {Jokerst},\ and\ \citenamefont {Padilla}}]{Liu2011}%
  \BibitemOpen
  \bibfield  {author} {\bibinfo {author} {\bibfnamefont {X.}~\bibnamefont
  {Liu}}, \bibinfo {author} {\bibfnamefont {T.}~\bibnamefont {Tyler}}, \bibinfo
  {author} {\bibfnamefont {T.}~\bibnamefont {Starr}}, \bibinfo {author}
  {\bibfnamefont {A.~F.}\ \bibnamefont {Starr}}, \bibinfo {author}
  {\bibfnamefont {N.~M.}\ \bibnamefont {Jokerst}}, \ and\ \bibinfo {author}
  {\bibfnamefont {W.~J.}\ \bibnamefont {Padilla}},\ }\href {\doibase
  10.1103/PhysRevLett.107.045901} {\bibfield  {journal} {\bibinfo  {journal}
  {Phys. Rev. Lett.}\ }\textbf {\bibinfo {volume} {107}},\ \bibinfo {pages}
  {045901} (\bibinfo {year} {2011})}\BibitemShut {NoStop}%
\bibitem [{\citenamefont {Lin}\ \emph {et~al.}(2003)\citenamefont {Lin},
  \citenamefont {Fleming},\ and\ \citenamefont {El-Kady}}]{Lin2003}%
  \BibitemOpen
  \bibfield  {author} {\bibinfo {author} {\bibfnamefont {S.-Y.}\ \bibnamefont
  {Lin}}, \bibinfo {author} {\bibfnamefont {J.~G.}\ \bibnamefont {Fleming}}, \
  and\ \bibinfo {author} {\bibfnamefont {I.}~\bibnamefont {El-Kady}},\ }\href
  {\doibase 10.1364/OL.28.001909} {\bibfield  {journal} {\bibinfo  {journal}
  {Opt. Lett.}\ }\textbf {\bibinfo {volume} {28}},\ \bibinfo {pages} {1909}
  (\bibinfo {year} {2003})}\BibitemShut {NoStop}%
\bibitem [{\citenamefont {Han}\ and\ \citenamefont {Norris}(2010)}]{Han2010}%
  \BibitemOpen
  \bibfield  {author} {\bibinfo {author} {\bibfnamefont {S.~E.}\ \bibnamefont
  {Han}}\ and\ \bibinfo {author} {\bibfnamefont {D.~J.}\ \bibnamefont
  {Norris}},\ }\href {\doibase 10.1103/PhysRevLett.104.043901} {\bibfield
  {journal} {\bibinfo  {journal} {Phys. Rev. Lett.}\ }\textbf {\bibinfo
  {volume} {104}},\ \bibinfo {pages} {043901} (\bibinfo {year}
  {2010})}\BibitemShut {NoStop}%
\bibitem [{\citenamefont {Vollmer}\ and\ \citenamefont
  {Arnold}(2008)}]{Vollmer2008}%
  \BibitemOpen
  \bibfield  {author} {\bibinfo {author} {\bibfnamefont {F.}~\bibnamefont
  {Vollmer}}\ and\ \bibinfo {author} {\bibfnamefont {S.}~\bibnamefont
  {Arnold}},\ }\href@noop {} {\bibfield  {journal} {\bibinfo  {journal} {Nat.
  Methods}\ }\textbf {\bibinfo {volume} {5}},\ \bibinfo {pages} {591} (\bibinfo
  {year} {2008})}\BibitemShut {NoStop}%
\bibitem [{\citenamefont {Sipe}\ and\ \citenamefont
  {Kranendonk}(1974)}]{Sipe1974}%
  \BibitemOpen
  \bibfield  {author} {\bibinfo {author} {\bibfnamefont {J.~E.}\ \bibnamefont
  {Sipe}}\ and\ \bibinfo {author} {\bibfnamefont {J.~V.}\ \bibnamefont
  {Kranendonk}},\ }\href {\doibase 10.1103/PhysRevA.9.1806} {\bibfield
  {journal} {\bibinfo  {journal} {Phys. Rev. A}\ }\textbf {\bibinfo {volume}
  {9}},\ \bibinfo {pages} {1806} (\bibinfo {year} {1974})}\BibitemShut
  {NoStop}%
\end{thebibliography}%

\end{document}